\documentclass[a4paper,11pt]{article}
\pdfoutput=1 % if your are submitting a pdflatex (i.e. if you have
\usepackage{jinstpub} % for details on the use of the package, please
\usepackage{subfigure}
\usepackage{siunitx}
\usepackage{amsmath}
\usepackage{graphicx}
\usepackage{xcolor}
\usepackage{multicol}
\usepackage{multirow}
\usepackage{url}

\title{\boldmath The performance of IHEP-NDL LGAD sensors after neutron irradiation}

\author[a,c]{Mengzhao Li}
\author[a,b]{Yunyun Fan}
\author[a,b]{Bo Liu}
\author[a,c]{Han Cui}
\author[a,c]{Xuewei Jia}
\author[a,c]{Shuqi Li}
\author[a,c]{Chengjun Yu}
\author[a,b]{Xuan Yang}
\author[a,b]{Wei Wang}
\author[a,c]{Mingjie Zhai}
\author[a,c]{Tao Yang}
\author[a,c]{Kewei Wu}
\author[a,c]{Yuhang Tan}
\author[a,c]{Suyu Xiao}
\author[a,b]{Mei Zhao} 
\author[a,b]{Xin Shi} 
\author[a,b]{Zhijun Liang}
\author[a,b,c]{Yuekun Heng}
\author[a,b]{Joao Guimaraes da Costa}
\author[d]{Xingan Zhang}
\author[d]{Dejun Han}
\author[e]{Alissa Howard}
\author[e]{Gregor Kramberger}

\affiliation[a]{Institute of High Energy Physics, Chinese Academy of Sciences, Beijing 100049, China}
\affiliation[b]{State Key Laboratory of Particle Detection and Electronics, Institute of High Energy Physics, Chinese Academy of Sciences, Beijing 100049}
\affiliation[c]{University of Chinese Academy of Sciences, Beijing 100049, China}
\affiliation[d]{Novel Device Laboratory, Beijing Normal University, Beijing 100875, China}
\affiliation[e]{Jozef Stefan Institute, SI-1000 Ljubljana, Slovenia}
\emailAdd{fanyy@ihep.ac.cn (Yunyun Fan)}
\emailAdd{liangzj@ihep.ac.cn (Zhijun Liang)}

\abstract{
The performances of Low Gain Avalanche diode (LGAD) sensors from a neutron irradiation campaign with fluences of \SI{0.8}{\times10^{15}}, \SI{1.5}{\times10^{15}} and \SI{2.5}{\times10^{15}~n_{eq}/\centi\metre^2} are reported in this article. These LGAD sensors are developed by the Institute of High Energy Physics, Chinese Academy of Sciences and the Novel Device Laboratory for the High Granularity Timing Detector of the High Luminosity Large Hadron Collider.
The timing resolution and collected charge of the LGAD sensors were measured with electrons from a beta source. 
After irradiation with a fluence of \SI{2.5}{\times10^{15}~n_{eq}/\centi\metre^2}, the collected charge decreases from 40 fC to 7 fC, the signal-to-noise ratio deteriorates from 48 to 12, and the timing resolution increases from 29 ps to 39 ps. 
}


\begin{document}
\maketitle
\flushbottom

%%%%
%%%% Introduction
%%%%
\section{Introduction}
\label{sec_intuduction}

The Low Gain Avalanche Diode (LGAD) is a new type of silicon sensor which has good timing resolution and moderate spatial resolution. 
This technology will be used to build the High Granularity Timing Detector (HGTD) for the ATLAS experiment in the High Luminosity Large Hadron Collider (HL-LHC)~\cite{HGTDtdr2020,HL-LHC,Nicolo2018}. 

\begin{figure}[htbp]
  \begin{center}
\rotatebox{0}{\includegraphics [scale=0.3]{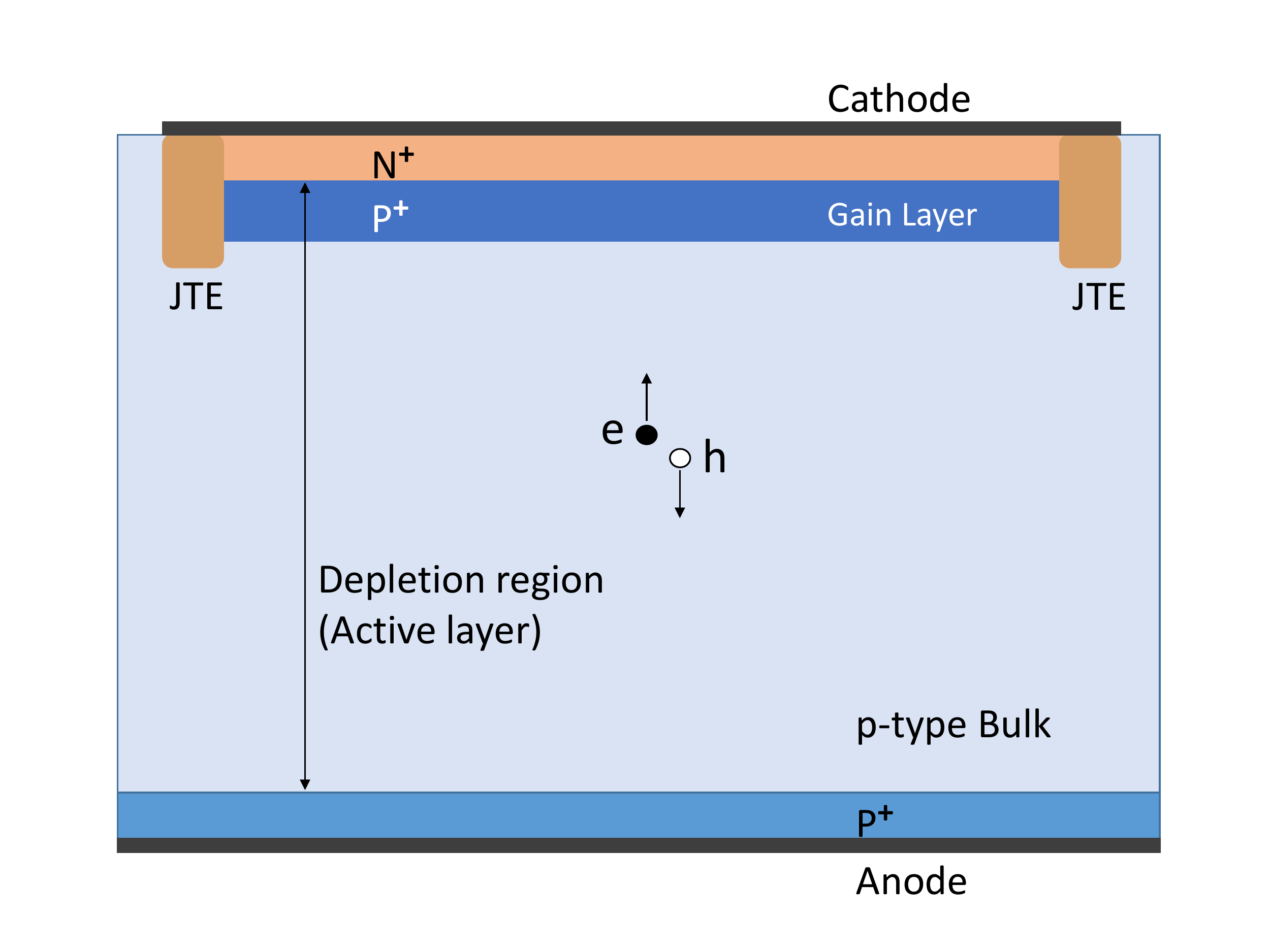}}% insert figure
\caption{Shematic for the LGAD sensors.}
\label{fig:SchematicLGAD}
 \end{center}
\end{figure}

The LGAD is a thin n-on-p silicon sensor with a highly doped $p^{+}$ region (as shown in Fig.~\ref{fig:SchematicLGAD}), where this $p^+$ region creates a high electric field and becomes a gain layer. 
Due to such a high electric field (> \SI{3}{\times10^{5}~V/cm}), more electron-hole pairs are generated by the initial carriers when passing through the gain layer. 
In the HGTD project, the timing resolution of LGAD sensor is required to be better than 70~ps after irradiation up to \SI{2.5}{\times10^{15}~n_{eq}/\centi\metre^2}. 
The collected charge with the same irradiation is required to be larger than 4~fC to reduce the jitter contribution in the readout ASIC chip~\cite{HGTDtdr2020}. 

The deteriorations of the collected charge and timing resolution for LGAD sensor after irradiation is due to the acceptor removal mechanism as described in Refs.~\cite{Moll2018,Ferrero2019,Kramberger2015,Tanyh2021}. 
Refs.~\cite{Cartiglia2017,Kramberger2018,Zhao2018,Lange2017,Shi2020,Wiehe2021} reported the radiation performance of HPK and CNM LGAD sensors.

%\note{HPK的Res和CC}
Previously, the Institute of High Energy Physics (IHEP) and the Novel Device Laboratory (NDL) jointly designed LGAD sensors with a \SI{33}{\micro\metre} active layer~\cite{Fan2020,NDL}. 
%(\note{somewhere you should mention active layer})
After proton irradiation with a fluence of \SI{1.5}{\times10^{15}~n_{eq}/\centi\metre^2}, the timing resolution can reach 45~ps with a collected charge of 2-4~fC. 
The performance of such LGAD sensor can not match the HGTD requirement for irradiation hardness (\SI{2.5}{\times10^{15}~n_{eq}/\centi\metre^2} irradiation).

To improve the radiation hardness, IHEP and NDL optimized the sensor design and fabricated a new type of LGAD sensor with an increased active layer thickness of \SI{50}{\micro\metre}.
%IHEP and IHEP-NDL also designed and produced a new type of 50 \SI{}{\micro\metre} active layer LGAD sensor. 
This article will present in detail the irradiation hardness performance of the new type \SI{50}{\micro\metre} LGAD after neutron irradiation up to \SI{2.5}{\times10^{15}~n_{eq}/\centi\metre^2}, including the capacitance, the leakage current, the collected charge, and the timing resolution. 
Besides, the factors affecting the time resolution after irradiation are analyzed in detail.

\begin{figure}[htbp]
\begin{center}
\includegraphics[scale=0.3]{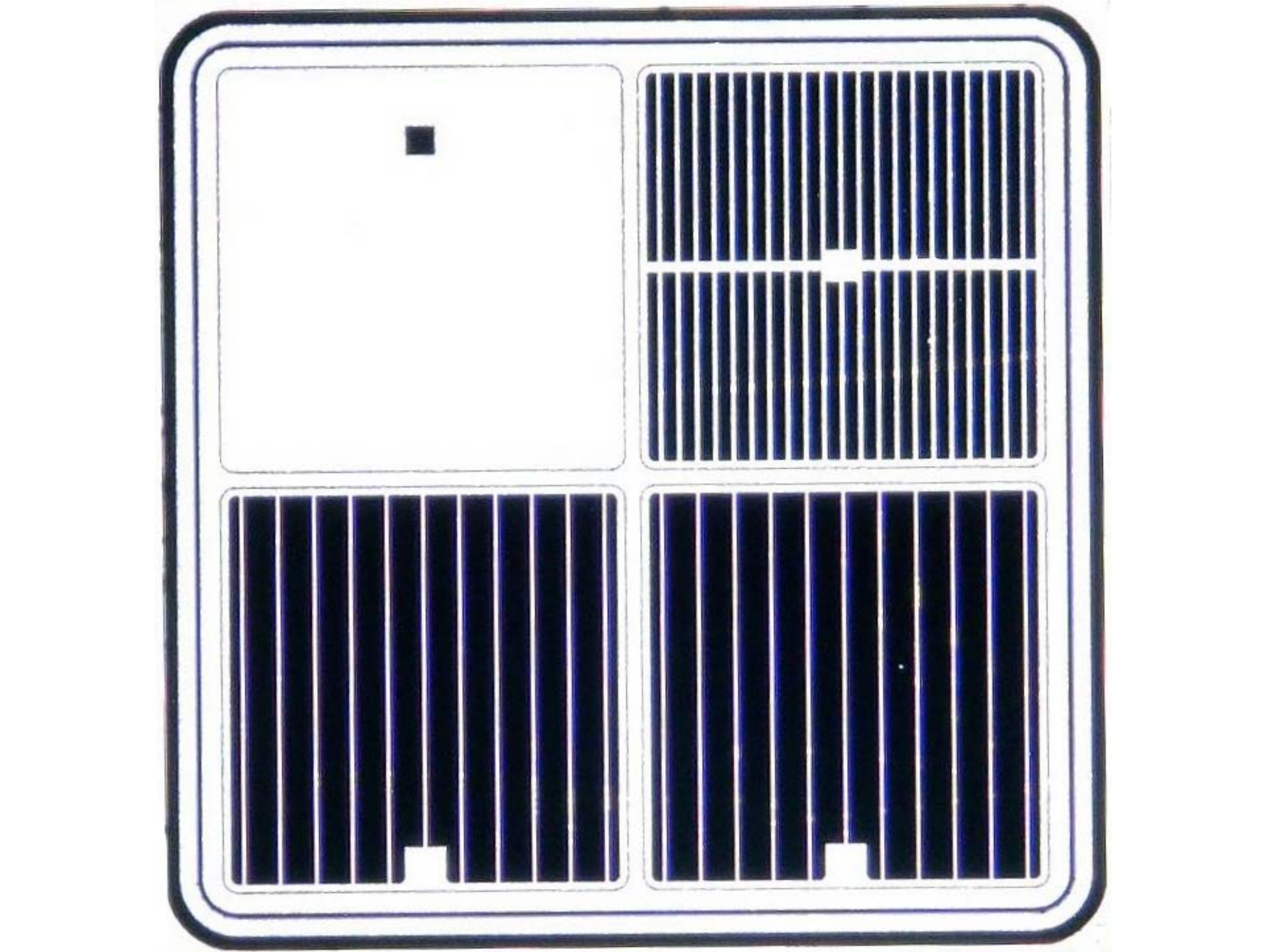} 
\caption{The layout of the \SI{50}{\micro\metre} LGAD sensors produced with single pad size of \SI{1.3}{mm} \si{\times} \SI{1.3}{mm}.}
\label{fig:Layout}
\end{center}
\end{figure}

%%%%
%%%% 2.Properties of IHEP-NDL 50 um LGAD
%%%%
\section{Properties of the IHEP-NDL \SI{50}{\micro\metre} LGAD}
\label{sec_Properties}

The IHEP-NDL \SI{50}{\micro\metre} LGAD was fabricated on a 6-inch wafer with a p-type epitaxial layer of a silicon substrate. 
The resistivity of the active layer is \SI{350}{~\Omega\cdot\centi\metre}. 
A highly-doped $p^{+}$ region is implemented to create a high electric field and moderate internal gain. 
The layout of IHEP-NDL \SI{50}{\micro\metre} LGAD shown in Fig.~\ref{fig:Layout}. 
It is 2~\si{\times}~2 arrays with three type electrodes, the single pad size is \SI{1.3}{mm}~\si{\times}~\SI{1.3}{mm}, and the lower left is a PIN without the gain layer. 
The upper left has a full coverage electrode, which is the electrode structure of the full-scale LGAD array.
%, \note{and is also the research object of this paper. What does this line mean?}
%The size of a single pad is 1.3 mm \si{\times} 1.3 mm, and the inter-pad gap is 40 \SI{}{\micro\metre}.

%%%%
%%%% 3.Neutron irradiation
%%%%
\section{Neutron irradiation}
\label{sec_irrad}

The \SI{50}{\micro\metre} LGADs were irradiated with neutrons by the Institut Jozef Stefan in Ljubljana, which has been used successfully in the past decades to support sensor development~\cite{Snoj2012}. 
In order to study the irradiation performance, three irradiation fluences are set, which are \SI{0.8}{\times10^{15}}, \SI{1.5}{\times10^{15}} and \SI{2.5}{\times10^{15}~n_{eq}/\centi\metre^2}. 
After irradiation, these sensors were annealed for 80 minutes at \SI{60}{\celsius}, which roughly emulates the long-term annealing at the end of year shut-down period during the operation of the HL-LHC. 
After that, sensors with three irradiation fluences were tested and studied.

%%%%
%%%% 4.IV-CV
%%%%
\section{Effect of irradiation on capacitance and leakage current characteristics}
\label{IV-CV}

Neutron irradiation affects the doping and the leakage current of the sensor.
The Capacitance-Voltage(C-V) and Current-Voltage(I-V) tests before and after irradiation are necessary.

%%%%
\subsection{Capacitance-Voltage(C-V)}

\begin{figure}[htbp]
\begin{center}
\subfigure[]{\includegraphics[width=.4\textwidth]{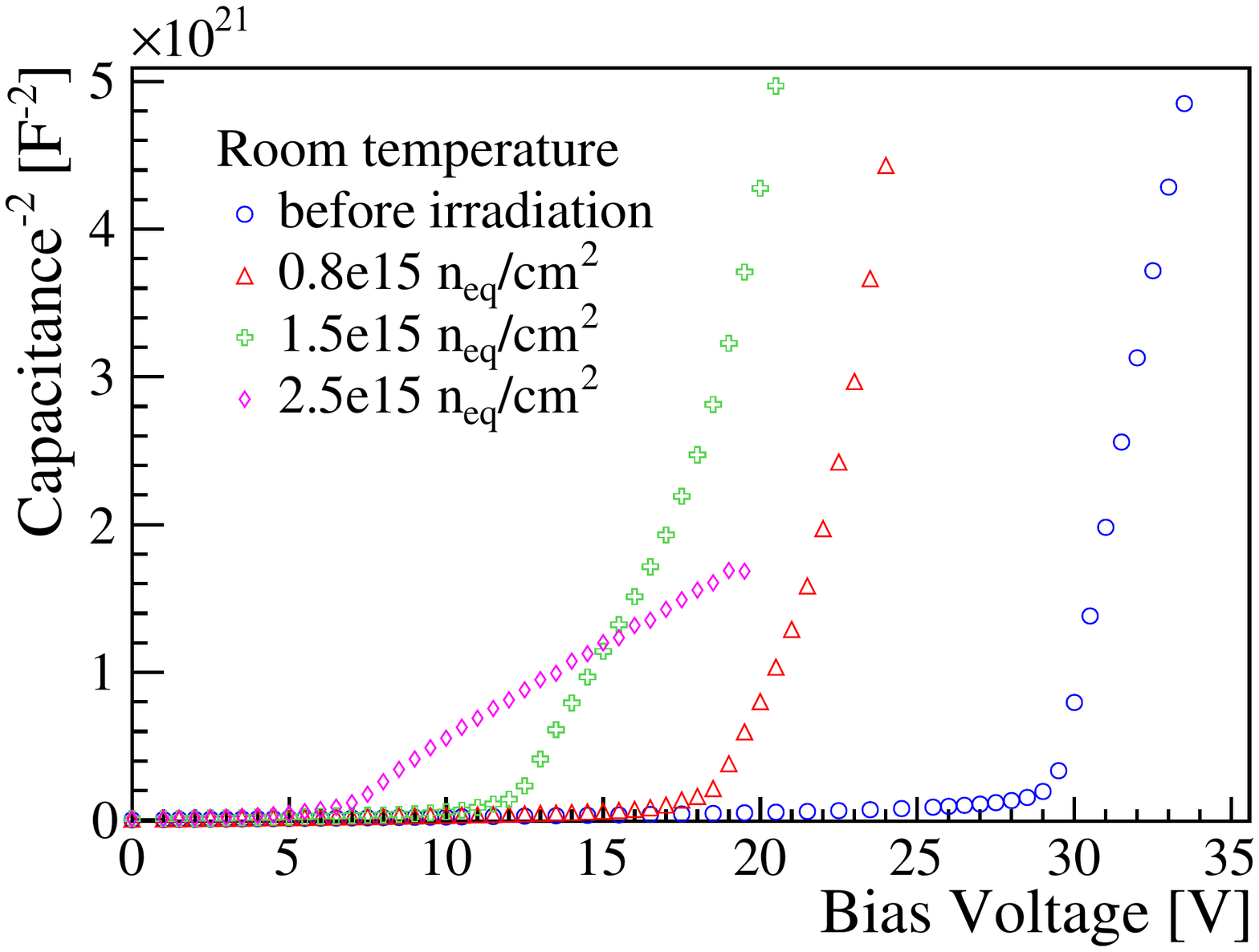} \label{fig:CVCurve}} 
\subfigure[]{\includegraphics[width=.4\textwidth]{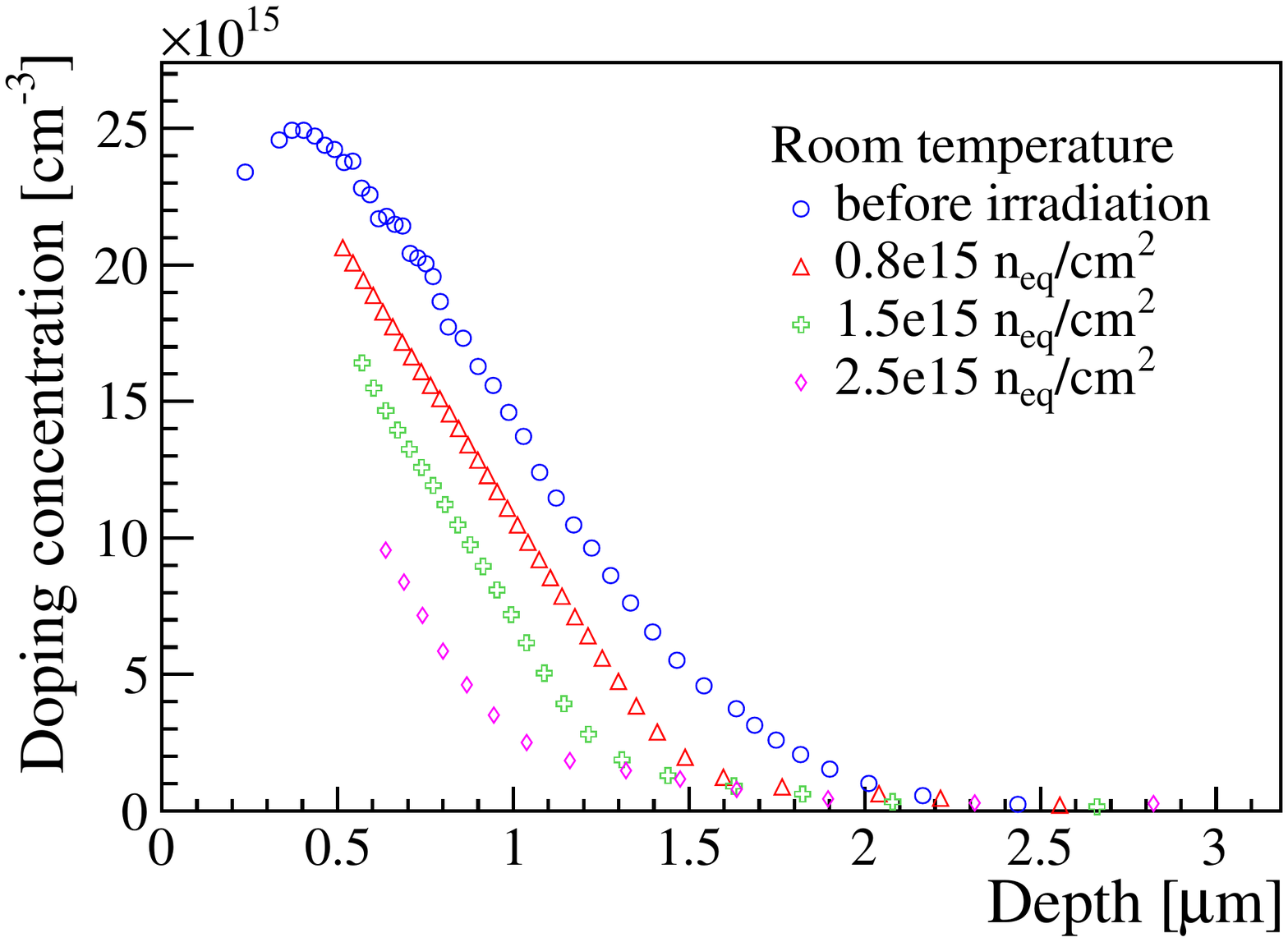} \label{fig:doping_depth}}
\caption{(a) ${1/C^{2}}$ as a function of the bias voltage for \SI{50}{\micro\metre} LGAD sensors at the room temperature. 
    (b) The doping profiles of \SI{50}{\micro\metre} LGAD sensors estimated from C-V measurements.
    }
\label{fig:CV_Dop}
\end{center}
\end{figure}

LCR meter was used to test the C-V characteristics of the \SI{50}{\micro\metre} LGAD sensors at room temperature. 
%(**The LCR meter frequency is set to 10 kHz for the sensor before the irradiation, and is set to 1 kHz for the sensor before the irradiation.) 
Fig.~\ref{fig:CVCurve} shows the inverse of capacitance square as a function of the bias voltage for sensors before and after the irradiation. 
The gain layer depletion voltage $V_{GL}$ of the sensor before the irradiation is 29 V, and drops to 18 V, 12 V, and 7 V with irradiation fluences of \SI{0.8}{\times10^{15}}, \SI{1.5}{\times10^{15}} and \SI{2.5}{\times10^{15}~n_{eq}/\centi\metre^2}, respectively.
With this C-V curve, the doping profile of the LGAD sensor was extracted ~\cite{Peiner1995,Nagai1992}. 
The relationship between doping concentration ${N_w}$ and depth ${w}$ is calculated with the following equations:

 \begin{equation}
     \begin{aligned}
	 N_{w} &= \frac{2}{{q\epsilon_{r}\epsilon_{0}A^2}}[\mathrm{d}(\frac{1}{C^2})/\mathrm{d}{V}]^{-1}, \\
	 w&=\epsilon_{r}\epsilon_{0}\frac{A}{C}.
     \end{aligned}
\end{equation}
where ${q}=\SI{1.6}{\times 10^{-19}~\coulomb}$ is the electron charge, $\epsilon_{r}=11.7$ is the relative permittivity of silicon, $\epsilon_{0}=\SI{8.854}{\times 10^{-12}}~\si{\mathrm{F}/\metre}$ is the permittivity of vacuum, ${A}$ is the area of the LGAD, ${V}$ is the bias voltage, and ${C}$ is the measured capacitance. 

The doping profiles for various irradiation conditions are shown in Fig.~\ref{fig:doping_depth}. 
Due to the acceptor removal mechanism~\cite{Ferrero2019}, the doping concentration of the gain layer decreases rapidly with the increase of the irradiation fluence. 
Compared with the case before irradiation, the doping concentration with the irradiation fluence of \SI{2.5}{\times10^{15}~n_{eq}/\centi\metre^2} reduces by a factor of 2.5. 
%This requires a higher bias voltage to compensate for the gain loss in the gain layer to obtain sufficient charge collection, \note{but the compensation is limited.}

\begin{figure}[htbp]
\begin{center}
\includegraphics[scale=0.4]{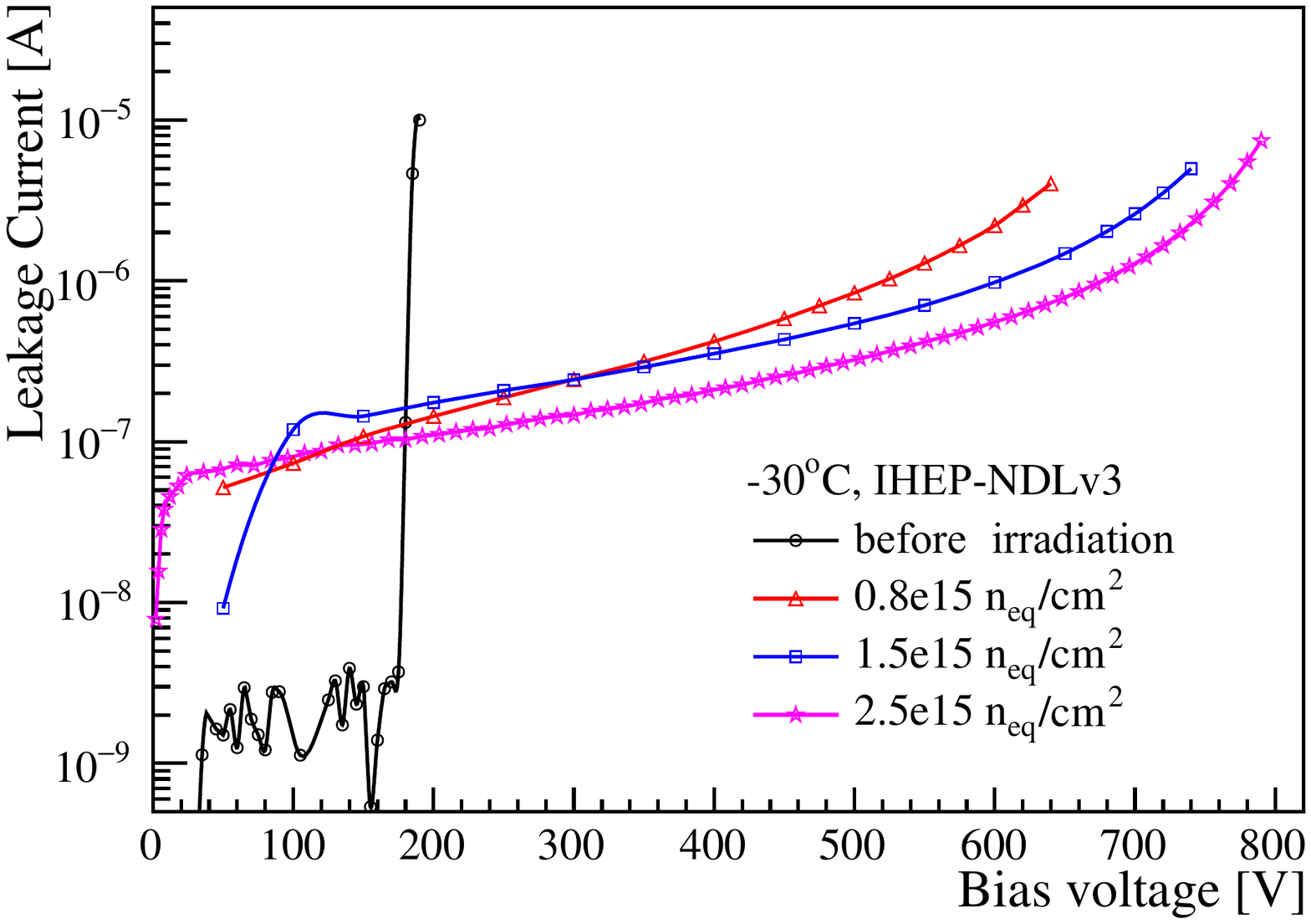} 
\caption{ I-V curves of 50 \SI{}{\micro\metre} LGAD sensors at the temperature of \SI{-30}{\celsius}.}
\label{fig:IV}
\end{center}
\end{figure}

%%%%
\subsection{Current-Voltage(I-V)}

Fig.~\ref{fig:IV} shows the leakage current of \SI{50}{\micro\metre} LGAD sensor in a dark environment at the temperature of \SI{-30}{\celsius}. 
The leakage current of the sensor before irradiation was at nA level, and its breakdowns at a voltage of \SI{190}{V}. 
When the irradiation fluence increases to \SI{2.5}{\times10^{15}~n_{eq}/\centi\metre^2}, the total leakage current (4 pads + GR) was \SI{3.36}{\micro\ampere} (\SI{50}{\micro\ampere/\centi\metre^2}) at the bias voltage of \SI{760}{V}. This is lower than the HGTD requirement(\SI{125}{\micro\ampere/\centi\metre^2}). The collected charge at such bias voltage reaches \SI{4}{fC}, which meets the HGTD requirements.

%%%%
%%%% 5.Low temperature beta telescope experiment
%%%%
\section{Low temperature beta telescope experiment}
\label{beta}
%%%%
\subsection{Beta telescope setup}

\begin{figure}[htbp]
\begin{center}
\includegraphics[scale=0.3]{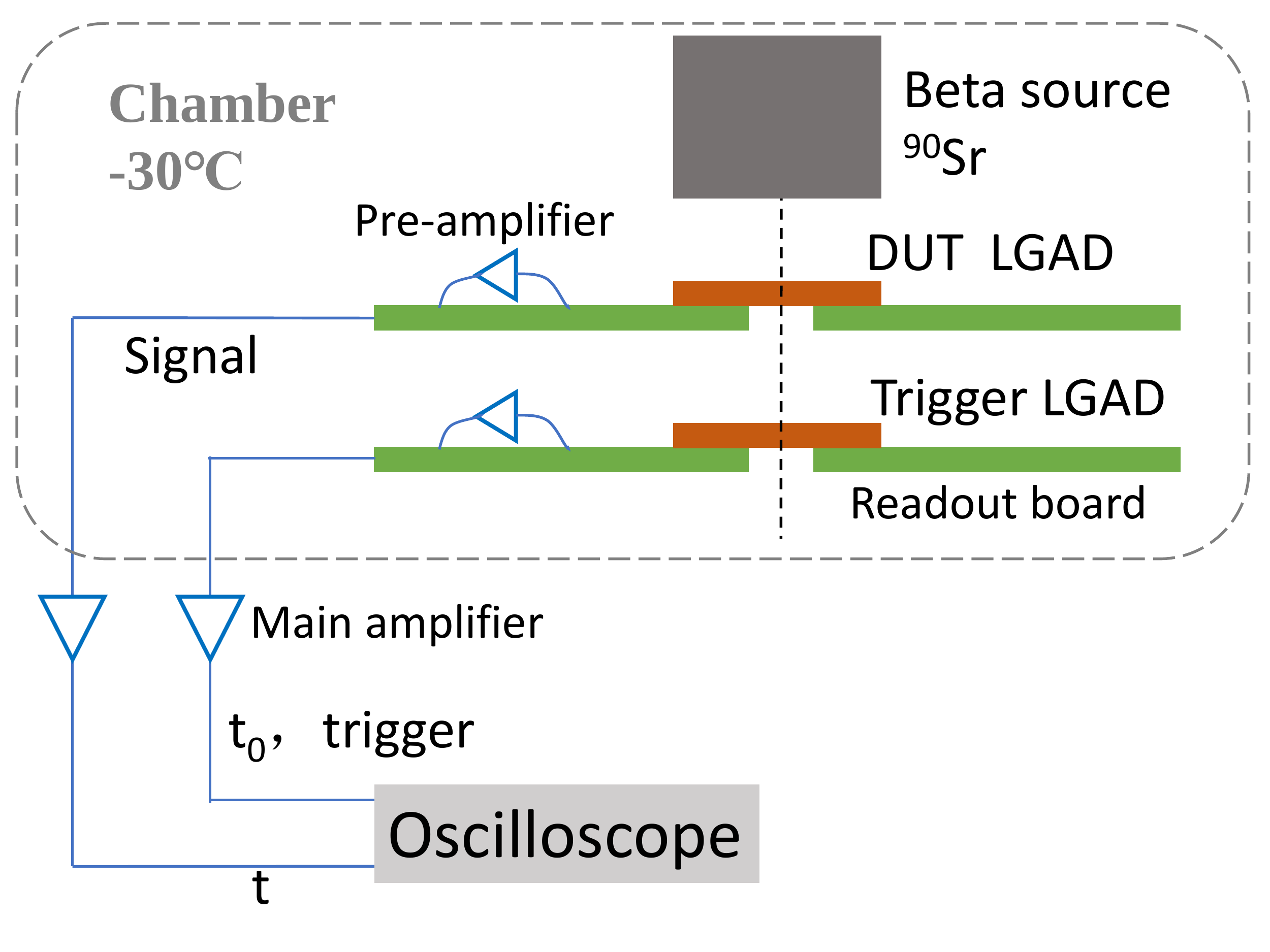} 
\caption{Beta telescope setup for IHEP-NDL sensors.}
\label{fig:BetaSetup}
\end{center}
\end{figure}

To study the timing resolution and the collected charge, LGAD sensors were tested with a beta source at the temperature of \SI{-30}{\celsius} with the humidity controlled within 10\%. 
The pad under test is wire bonded to the readout board, while the guard ring and the other three pads are grounded. 
The readout board is a single channel readout board designed by the University of California Santa Cruz (UCSC)~\cite{Cartiglia2017}. The readout board has a high-speed inverting preamplifier with a trans-impedance of \SI{470}{~\Omega}. 
This preamplifier is followed by an external main amplifier with a gain of \SI{20}{~dB}. 
Fig.~\ref{fig:BetaSetup} shows the telescope experiment setup with a beta source. The top sensor is the device under test (DUT), and the bottom sensor is used to trigger electron signals from the beta source. 
The signal pulses from both sensors are recorded by a digital oscilloscope for offline analysis.

\begin{figure}[htbp]
\begin{center}
\includegraphics[scale=0.4]{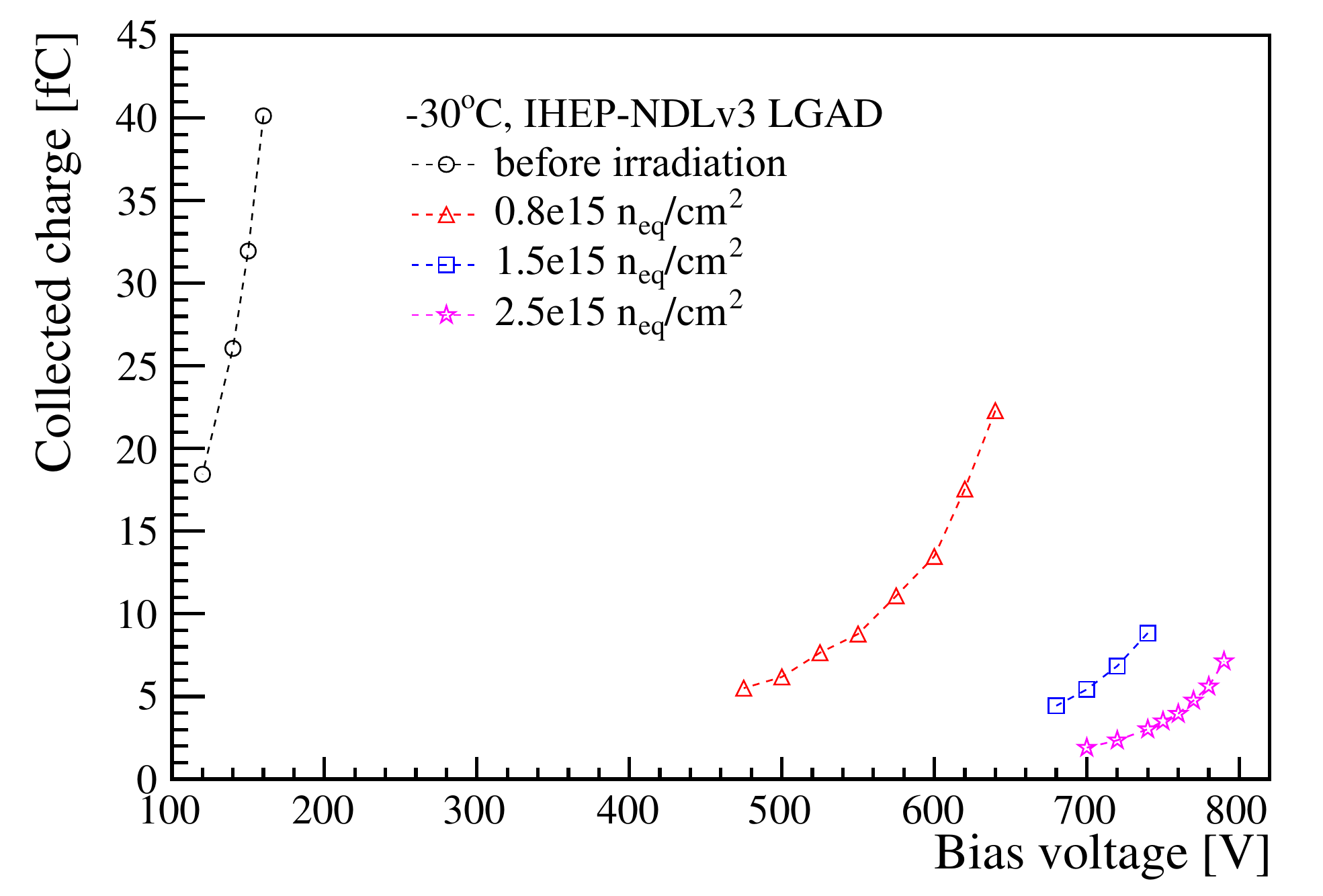} 
\caption{The collected charge as a function of bias voltage for IHEP-NDL \SI{50}{\micro\metre} LGAD before and after irradiation at temperature of \SI{-30}{\celsius}. }
\label{fig:Charge}
\end{center}
\end{figure}

%%%%
\subsection{Collected charge}
\label{CC}

The collected charge for each sensor is calculated as the integration of the pulse shape divided by the total gain of the preamplifier and the main amplifier. 
%Note that the overshoot of the pulse was placed outside the integration range. 
The distribution of the collected charge is fit with a shape from the convolution of the Landau and the Gaussian functions to get the most probability value (MPV) for the collected charge. 

The collected charges of the \SI{50}{\micro\metre} LGAD sensor before and after irradiation are shown in Fig.~\ref{fig:Charge}. 
The LGAD sensor before irradiation has a very high collected charge of \SI{40}{fC}.  
When the irradiation fluence increased to \SI{2.5}{\times10^{15}~n_{eq}/\centi\metre^2}, the collected charge of the \SI{50}{\micro\metre} LGAD sensor reaches \SI{4}{fC} at a bias voltage of \SI{760}{V}, and keep increasing with the higher bias voltage to \SI{7}{fC} for current measurement. 
As the neutron fluence increasing, the space charge density of the gain layer decreases due to the acceptor removal mechanism. 
Even with a high bias voltage, the ability of the sensor to collect charge decreases.
%This requires a higher bias voltage to compensate for the gain loss in the gain layer to obtain sufficient charge collection, \note{but the compensation is limited.}
This result indicates that the IHEP-NDL \SI{50}{\micro\metre} sensor passes the HGTD requirement for charge collection with irradiation (\textgreater \SI{4}{fC} after \SI{2.5}{\times10^{15}~n_{eq}/\centi\metre^2} irradiation fluence).

%%%%
\subsection{Timing resolution}
\label{Timing}

The timing resolution $\sigma_{t}$ of charge particles passing through LGAD sensors is mainly composed of the following three items: %contributions from TimeWalk, from Landau fluctuation and from Jitter. 
\begin{equation} \label{eq:Timing}
\sigma^2_t = \sigma^2_{\mathrm{TimeWalk}} + \sigma^2_{\mathrm{Landau}} +  \sigma^2_{\mathrm{Jitter}} 
%\label{eq:sigmat}
\end{equation}

The time walk effect ($\sigma_{\mathrm{TimeWalk}}$) is caused by different signal amplitudes, which can be corrected by using the constant fraction discriminator (CFD) method. The details of the CFD method are in reference \cite{Cartiglia2017}.
The jitter effect ($\sigma_{\mathrm{Jitter}}$) is proportional to the noise and the inverse of the signal slope~\cite{Nicolo2018}:

\begin{equation} \label{eq:Jitter}
\sigma^2_{\mathrm{Jitter}} = \left(\frac{N}{\mathrm{d}V/\mathrm{d}t}\right)^2, %\simeq \frac{t_{\mathrm{rise}}}{(S/N)}.
\end{equation}

where ${N}$ is the Root Mean Square (RMS) of noise, d${V/}$d${t}$ is the slope of the rising edge around the threshold value.
A better signal-to-noise ratio (SNR) introduces a smaller jitter contribution to the timing resolution. 
The Landau contribution ($\sigma_{\mathrm{Landau}}$) corresponds to the effect from Landau fluctuations in the energy deposition non-uniformities along the path of a particle passing through the detector. 
%Among them, $\sigma_{TimeWalk}$ is the time walk effect caused by different signal amplitudes, which can be corrected by the constant fraction discriminator (CFD) method; $\sigma_{Landau}$ is caused by the non-uniform energy deposition; $\sigma_{Jitter}$ is proportional to the noise RMS and inversely proportional to the slope of the rising edge. 

The CFD method is used to calculate the flight time ($\Delta{t}$) of the electron between two LGAD sensors. 
The detailed calculation for the timing resolution is described in Ref.~\cite{Cartiglia2017}. 
Since the timing resolution of the trigger sensor is known, the timing resolution of the DUT sensor is derived as:

\begin{equation}
\sigma_{\mathrm{DUT}} = \sqrt{\sigma^2_{\Delta{t}} - \sigma^2_{\mathrm{Trigger}}} 
%\label{eq:thetaT}
\end{equation}

\begin{figure}[htbp]
\begin{center}
\includegraphics[scale=0.4]{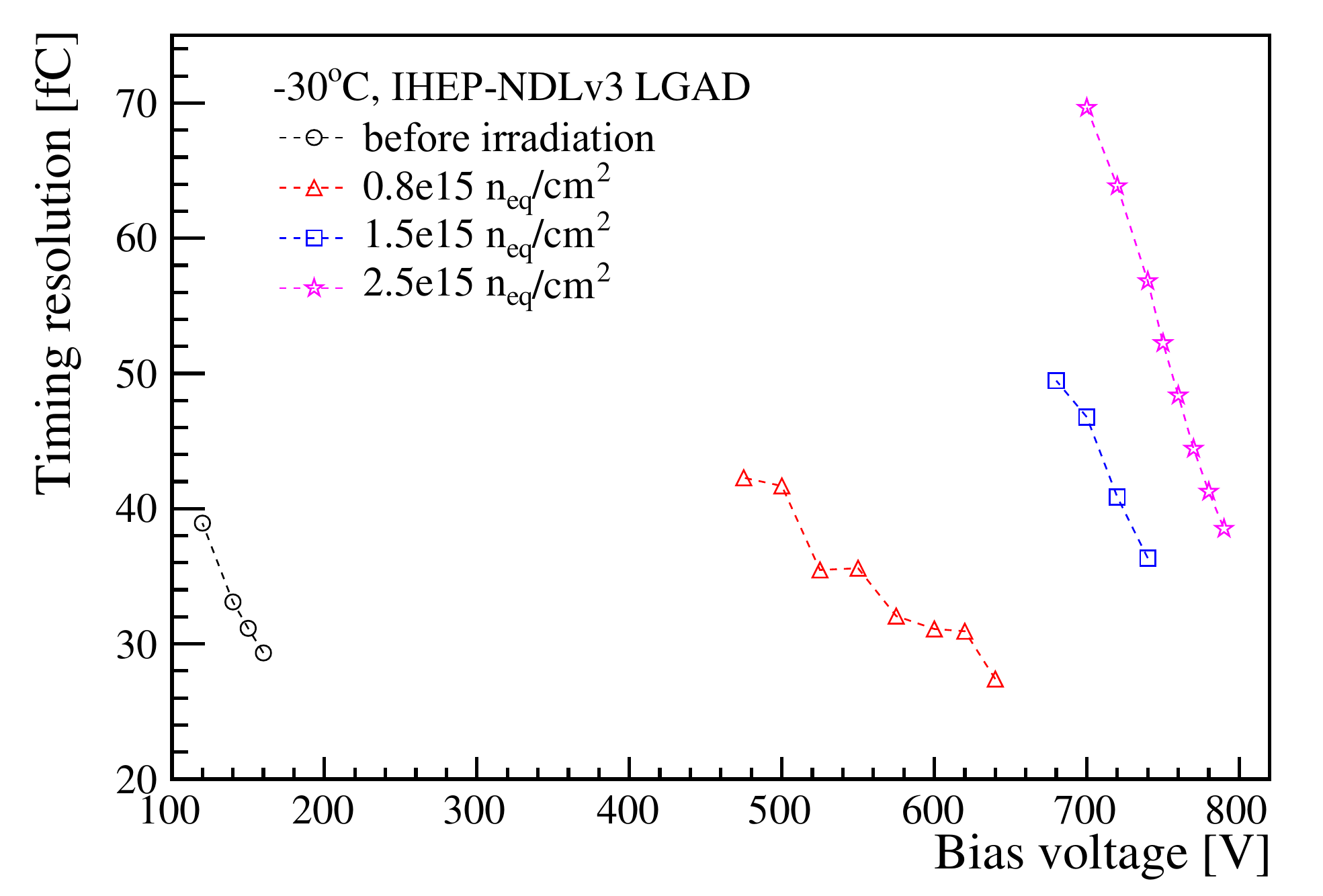} 
\caption{Timing resolution as a function of bias voltage for IHEP-NDL \SI{50}{\micro\metre} LGAD sensor before and after irradiation at temperature of \SI{-30}{\celsius}.}
\label{fig:Res}
\end{center}
\end{figure}

The timing resolution of the \SI{50}{\micro\metre} LGAD sensors before and after the irradiation is shown in Fig.~\ref{fig:Res}. 
The timing resolution before the irradiation can reach \SI{26}{ps}. 
As the increase of the irradiation fluence, the working voltage increases quickly.
To have a same timing resolution of \SI{40}{ps}, the sensor irradiated with three different fluences needs bias voltages of \SI{500}{V}, \SI{720}{V} and \SI{780}{V} respectively. 
%Timing resolutions of sensor with three irradiation fluences reach \SI{50}{ps} at \SI{450}{V}, \SI{680}{V}, and \SI{760}{V} respectively. 
The IHEP-NDL \SI{50}{\micro\metre} LGAD sensor meets the HGTD requirement for the timing resolution (\textless \SI{70}{ps} after \SI{2.5}{\times10^{15}~n_{eq}/\centi\metre^2} irradiation fluence).

%%%%
\subsection{Jitter and Landau contributions}

The irradiation increases the working voltage, reduces the collected charge, and deteriorates the timing resolution. The Jitter and the Landau contributions are the main sources for the timing resolution.
%Contributions of the Jitter and the Landau in the timing resolution with the change of radiation dose needs further analysis.
Their contributions in the timing resolution as various irradiation fluences are analyzed and presented in this section.

\begin{figure}[htbp]
\begin{center}
\subfigure[]{\includegraphics[width=.6\textwidth]{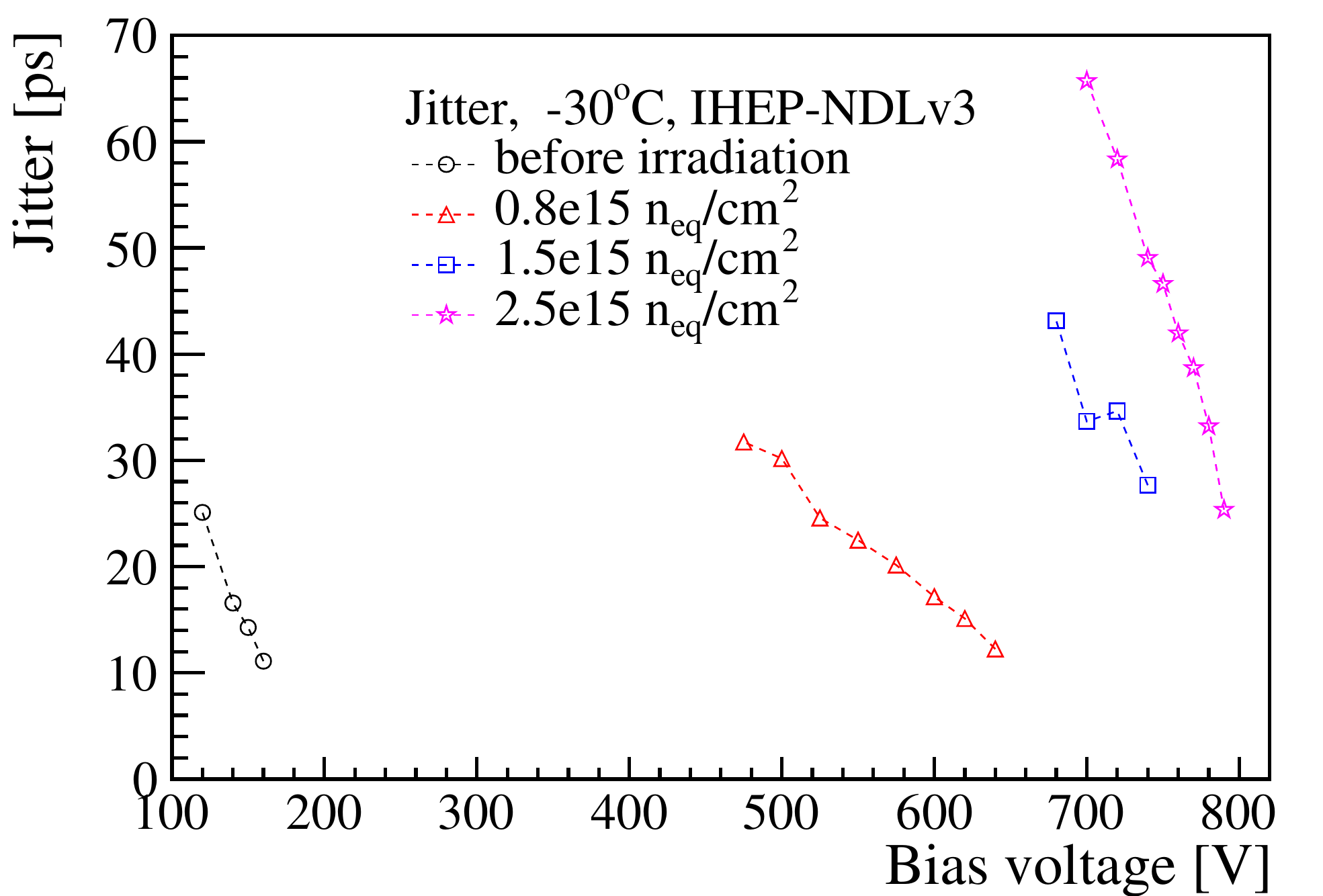} \label{fig:Jitter_BV} }
\subfigure[]{\includegraphics[width=.6\textwidth]{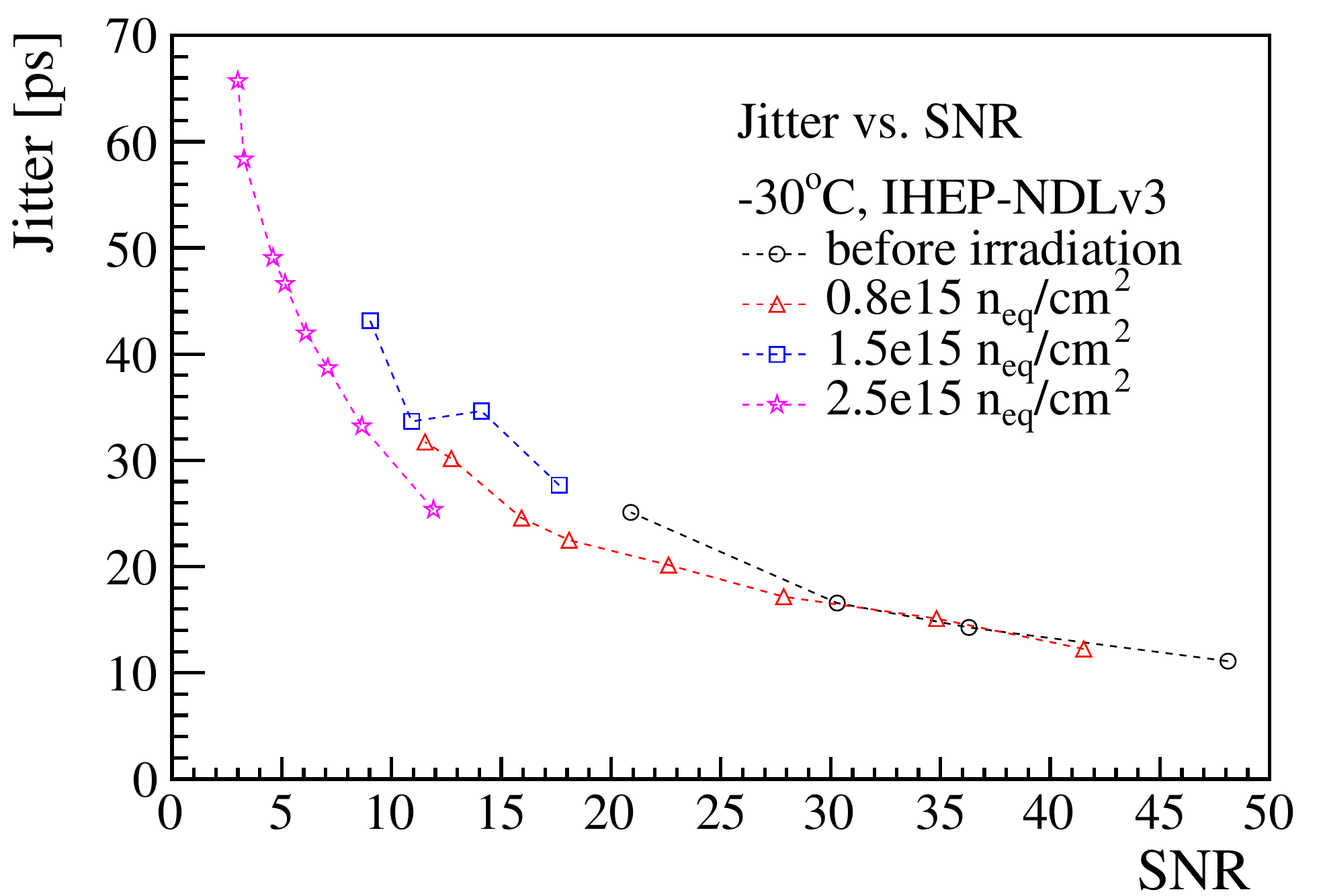}\label{fig:Jitter_SNR} }
\subfigure[]{\includegraphics[width=.6\textwidth]{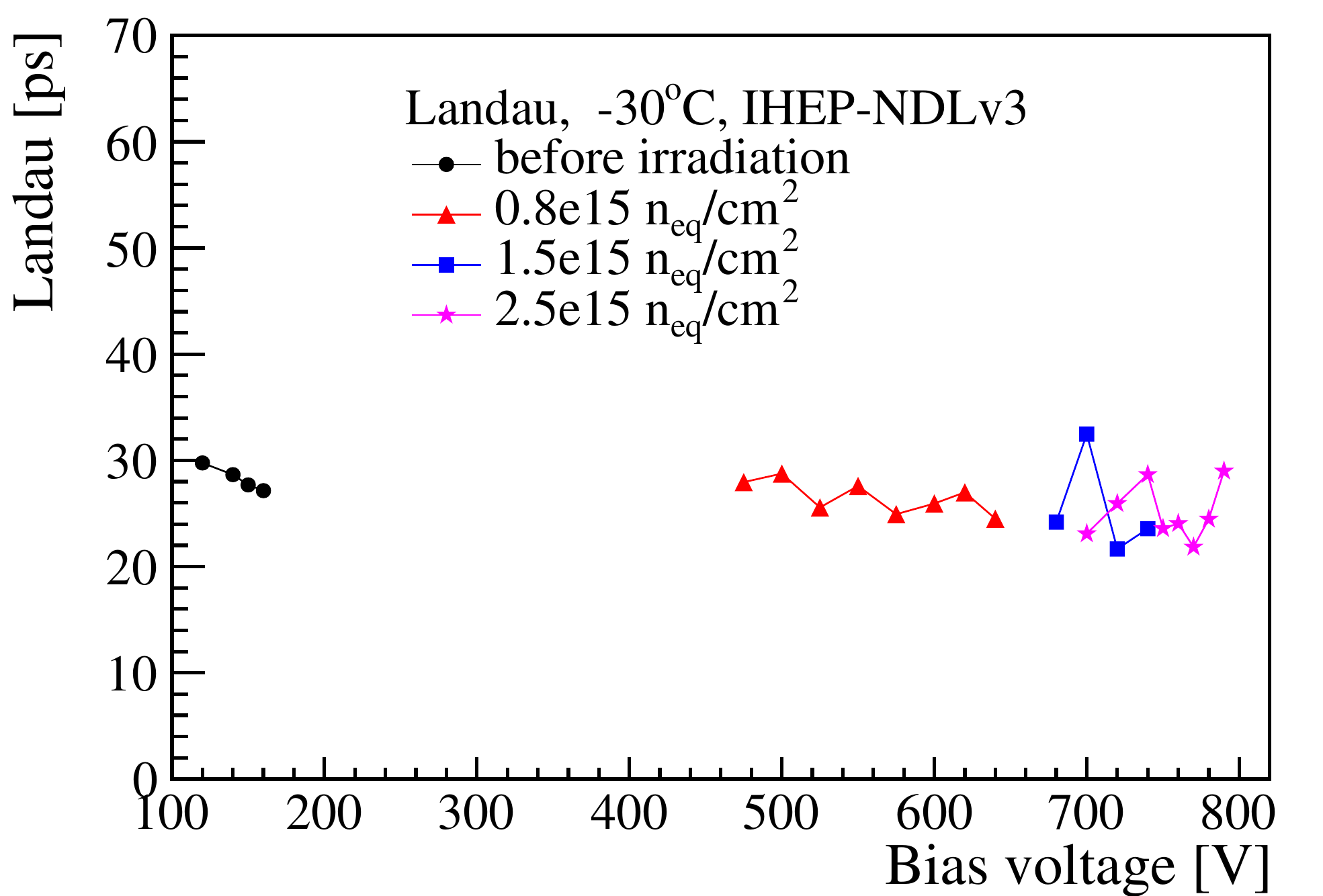} \label{fig:Landau_BV} }
\caption{(a) The Jitter contribution as a function of the bias voltage for \SI{50}{\micro\metre} IHEP-NDL sensors. 
(b) The Jitter contribution as a function of the signal-to-noise ratio for IHEP-NDL sensors. 
(c) The Landau contribution as a function of the bias voltage for IHEP-NDL 50 \SI{}{\micro\metre} LGAD before and after the irradiation at the temperature of \SI{-30}{\celsius}. }
\label{fig:Jitter_landau}
\end{center}
\end{figure}

The Jitter contribution is calculated with Eq.~(\ref{eq:Jitter}). 
Fig.~\ref{fig:Jitter_BV} shows the Jitter contributions with different irradiation fluences as functions of the bias voltage and the SNR. 
The Jitter contribution decreases with the bias voltage increase and increases with the irradiation fluence. 
As the irradiation fluence increasing, acceptors in the gain layer are quickly removed, resulting in a gain decrease. 
This decrease of the gain results in a worse SNR and a larger Jitter contribution.  
As shown in Fig.~\ref{fig:Jitter_SNR}, The Jitter and SNR have a negative index relationship.  
When the SNR is less than 10, the Jitter contribution increases rapidly.
The SNR of the IHEP-NDL \SI{50}{\micro\metre} sensor before the irradiation can be as high as 48, and reduces by a factor of 4 to be 12 after the irradiation with a fluence of \SI{2.5}{\times10^{15}~n_{eq}/\centi\metre^2}.

The Landau contribution is calculated from Eq.~(\ref{eq:Timing}). 
Fig.~\ref{fig:Landau_BV} gives the Landau contribution for different irradiation fluences. 
The Landau contribution is about 25-30 ps, and does not change dramatically with the bias voltage and the irradiation fluence. 
The reason is that the Landau contribution is caused by the non-uniform energy deposition and  
the bias voltage and the irradiation do not significantly affect this non-uniform energy deposition of the electron in the epitaxial layer.

%%%%
%%%% 6.Conclusion
%%%%
\section{Conclusion}
\label{sec_Conclusion}

The new \SI{50}{\micro\metre} LGAD sensor was designed by the IHEP and the NDL. 
It was irradiated by neutrons and tested with the beta telescope at the temperature of \SI{-30}{\celsius}. 
The doping concentration of the gain layer drops quickly after irradiation, and a higher working voltage is required to compensate for the gain. 
After the irradiation with a fluence of \SI{2.5}{\times10^{15}~n_{eq}/\centi\metre^2}, the collected charge meets the HGTD requirements (4 fC) at the bias voltage of \SI{760}{V}. 
Besides, the leakage current is \SI{50}{\micro\ampere/\centi\metre^2} at the same bias voltage, which also passes the HGTD requirements (125 \SI{}{\micro\ampere/\centi\metre^2}). 
The Landau contribution is almost stable with different irradiation fluences and bias voltages as expected. 
The decreases of the gain and the SNR for the irradiated sensor lead to an increase of the jitter contribution, and eventually worsen the timing resolution. 
After irradiation  with a fluence of \SI{2.5}{\times10^{15}~n_{eq}/\centi\metre^2}, the timing resolution can still reach \SI{39}{ps}, which meets the HGTD requirements (70 ps). 
The HEP-NDL \SI{50}{\micro\metre} LGAD sensor performs a very good irradiation hardness and has the potential to be used in the HGTD project for the ATLAS detector upgrade during the HL-LHC. 
%In the next prototype run, IHEP and NDL will optimize the p+ region doping to reduce the working voltage for the irradiated sensor. (\note{maybe remove this line?})

%\appendix
%\section{Some title}
%Please always give a title also for appendices.

\acknowledgments

This work was supported by the National Natural Science Foundation of China (No.11961141014), the State Key Laboratory of Particle Detection and Electronics (SKLPDE-ZZ-202001), the Hundred Talent Program of the Chinese Academy of Sciences (Y6291150K2), the CAS Center for Excellence in Particle Physics (CCEPP). Thanks to Beijing Normal University for the detector production.

%\paragraph{Note added.} This is also a good position for notes added
%after the paper has been written.


\end{document}